\documentclass[journal]{IEEEtran}

\usepackage{amssymb} 
\usepackage[short]{optidef} 
\usepackage{eqnarray}
\usepackage{algorithm}
\usepackage{mathtools}
\usepackage{amssymb}
\usepackage{amsfonts}
\usepackage{amsmath}
\usepackage{amsthm}
\usepackage{algorithm}
\usepackage{algorithmic}
\usepackage{blindtext}
\usepackage{ifpdf}
\usepackage{graphicx}
\usepackage{cite}
\usepackage{color,soul}
\usepackage{xcolor}
\usepackage{balance}
\usepackage{setspace}
\usepackage{float} 
\usepackage{mathtools, nccmath}

\usepackage{caption}

\newtheorem{remark}{Remark}

\newcommand{\diag}{\text{diag}}
\newcommand{\cov}{\text{cov}}

\title{Two-level Robust State Estimation for Multi-Area Power Systems  Under Bounded Uncertainties}

\author{Shiva Moshtagh and Mehdi Rahmani
\thanks{Shiva Moshtagh and Mehdi Rahmani are with Department of Electrical Engineering, Imam-Khomeini International University, Qazvin, Iran (email: mrahmani@eng.ikiu.ac.ir)}}
\begin{document}

\maketitle

\begin{abstract}
This paper introduces a two-level robust approach to estimate the unknown states of a large-scale power system while the measurements and network parameters are subjected to uncertainties. The bounded data uncertainty (BDU) considered in the power network is a structured uncertainty which is inevitable in practical systems due to error in transmission lines, inaccurate modelling, unmodeled dynamics, parameter variations, and other various reasons. In the proposed approach, the corresponding network is first decomposed into smaller subsystems (areas), and then a two-level algorithm is presented for state estimation. In this algorithm, at the first level, each area uses a weighted least squares (WLS) technique to estimate its own states based on a robust hybrid estimation utilizing phasor measurement units (PMUs), and at the second level, the central coordinator processes all the results from the subareas and gives a robust estimation of the entire system. The simulation results for IEEE 30--bus test system verifies the accuracy and performance of the proposed multi-area robust estimator. 
\end{abstract}

\begin{IEEEkeywords}
State estimation, weighted least squares, phasor measurement unit, uncertainty, multi-area power system
\end{IEEEkeywords}


\section{Introduction}
\IEEEPARstart{S}{tate} estimation (SE) determines the most likely states of a power system using the measurements collected by SCADA system via remote terminal units (RTUs). Since measurements are not available exactly, the state variables are estimated based on the statistical methods. In this regard, weighted least squares (WLS) is one of the most common criteria that minimizes the sum of the squares of weighted differences between the estimated and actual variables. As the current electrical power systems continue to become larger and more complex, there has been a rising demand for computationally efficient robust SE algorithms. Also, to complete the process of estimation, information is collected from smaller and often competing companies. These companies usually employ distinctive algorithms for SE, and may be reluctant to share the detailed information with their rivals. Therefore, this estimation is performed in a distributed paradigm for multi-area power systems.

Research on multi-area SE can be traced back to the 1970s \cite{r1}, right after the introduction of SE in power systems by Schweppe \cite{schweppe1}. In multi-area SE, the network should first be broken down into individual areas. There are different strategies for network decomposition. In \cite{falcao1995parallel,muscas2015multiarea}, the system is decomposed such that the areas (subsystems) are overlapping; while in \cite{r5,le2019,rathod2019}, they are non-overlapping. The latter kind of decomposition can speed up and simplify the estimation process because the estimation is separately performed at each area in a parallel manner. Most of multi-area SE techniques utilize a coordinator to handle the interaction of local estimators \cite{c2,c1,c3,c5}; however, some others 
do not require a coordinator \cite{h1,li2020}. In this regard, a two-stage multi-area approach is presented in \cite{r8} in which at the first stage, local estimators at each area operate separately. The second stage integrates the solutions from all local estimators along with other available measurements to give a central estimation of the whole system via coordinator. In \cite{taxonomy}, a taxonomy of multi-area SE has been proposed based on multiple classification criteria. 

In the recent years, the advent of phasor measurement unit (PMU) has revolutionized power system SE due to their high speed and accuracy. Before these digital devices, SE relied on the measurements obtained from a SCADA system which includes traditional measurements such as bus power injections and line power flows. It is shown that with the aid of enough PMUs, the performance of multi-area SE will improve \cite{r8}. However, Since it is not economically and technically possible to have enough number of PMUs in a large-scale network, the optimal PMU placement has also been a favorite research topic recently \cite{c2}. By using only PMU measurements, a linear SE can be utilized to determine all states of a observable system. In \cite{xu2016,xu2017}, a multi-area robust estimation has been addressed using only PMUs through a linear model. PMUs are generally utilized along with the traditional measurements. This scheme is so-called \textit{hybrid} estimation method. This combination can either be directly \cite{r26}, or indirectly \cite{r27}. As a two-step hybrid method, \cite{zhou2006} proposed an algorithm in which a traditional SE is first executed via SCADA measurements, then the results of the previous step as well as PMU measurements will be incorporated in a linear post processing estimator. 

It is very important to note that error in transmission of information through the network, inaccurate modeling, etc cause uncertainty in the network model. It can affect the accuracy of the estimation significantly \cite{IEEE5}. Since the WLS method is sensitive to inaccurate data, it is not a robust method for estimation. In the literature, several methods have been studied towards developing robust alternatives to this non-robust estimator. Among all, the least absolute value (LAV) estimator could be a practical and more robust approach providing that there are not leverage measurements \cite{abur1991}. In \cite{gol2014}, the problem of leverage measurements has been solved by using only PMU measurements which could not be affordable. Several multi-area robust approaches have been provided in \cite{kekatos2012,li2013,lin2019} by handling bad data related to the outlier measurements which deteriorate the performance of the estimator.
Model validation can also be efficient along to safeguarding the estimation process against uncertainty, yet is challenging \cite{qi2018}. Accordingly, it is a more viable way to make estimators more robust to model uncertainty. As far as we know, the robust SE problem for power systems subject to norm-bounded uncertainties using PMUs has not been studied in the literature yet, in particular, in a distributed (two-level) approach for multi-area power systems.

This paper presents a novel algorithm for multi-area robust SE to estimate the states of an uncertain large-scale power system using SCADA and PMU measurements. The contribution of this paper is primarily considering bounded structured uncertainty in both measurements and network parameters of the entire power system and then, implementing a multi-area estimation through a two-level approach integrating hybrid linear method. Initially, the entire system is decomposed into a specific number of non-overlapping areas. Then, at the first level of the algorithm, the individual areas separately perform their estimators in parallel based on their internal measurements. The central coordinator marks the second level of the estimation process by receiving estimated values of states and boundary measurements as well, and finally gives a robust SE for the power system. It is remarkable that this idea has not been applied to power systems' SE so far. The reason for employing such a technique arises from firstly, the accuracy and secondly, the fast speed of the proposed distributed algorithm which does not require iteration between the local estimators and the coordinator. This multi-area algorithm does not enforce any particular requirements on the boundary measurements. Moreover, each area is empowered to have its own estimation algorithm, network database, and measurement without affecting the performance of the other areas' estimators since the local estimators do not exchange or interact data. It is assumed that there is enough redundancy to ascertain the observability, and bad data have been eliminated; however, bounded data uncertainty as another major source of misestimating is imposed on the measurements.

The remainder of this paper is structured as follows. Section \ref{problem} introduces the formulation of multi-area SE. Section \ref{section3} addresses a brief review on WLS and robust estimation separately. Section \ref{framework} focuses on a framework which presents the algorithm and structure of the proposed method. In Section \ref{simulation} simulation results on IEEE 30--bus test system are discussed and in Section \ref{conclusion} conclusion remarks are presented.\\

\noindent
\textbf{Notations:} The notations used throughout the paper are fairly standard which are explained as follows

\begin{table}[H]
\centering
\caption*{}
\scalebox{1}{
\begin{tabular}{c|c c c}
Notation & Meaning  \\
\hline
\hline
 $A^{-1}$ & Inverse of matrix A \\

 $A^{\dag}$ & Moore-Penrose pseudoinverse of matrix A \\

 $\mathbb{E}(a)$ & Mathematical expectation of variable a \\

 $\cov(A)$ & Covariance of matrix A \\

 $\diag(A)$ & Diagonal elements of matrix A \\

 $\| a \|$ & Norm of variable a \\

 $ \mathbb{R}^+ $ & Coordinate space over the positive numbers \\

\end{tabular}}
\end{table}

\section{Multi-area SE Modeling} \label{problem}
In the proposed multi-area SE, the large-scale power network should be decomposed into non-overlapping local areas. The partitioned zones are connected to each other by tie-lines. The state variables in a power system are bus voltage magnitude and phase angle that are needed to be estimated through some measurements including bus power injections, and line power flows.

For multi-area decomposition, a power network consisting of $n$ buses is considered to be decomposed into $r$ areas. This division is done in such a way that the minimum boundary tie-lines exist in the decomposed network. Accordingly, at $i$th area, each bus can be one of these types:
\begin{itemize}
	\item Internal bus: buses that are completely within area $i$ and their neighbors are members of the same area.
	\item Boundary bus: buses that have at least one connection to buses from other areas.
	\item External bus: buses that are located at another area and have at least one link to a boundary bus at area $i$.
\end{itemize}

\begin{figure}
  \centering
  \includegraphics[scale=0.7]{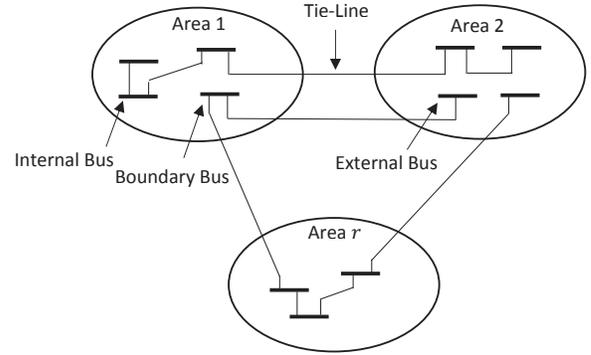}
  \caption{Decomposed power network in $r$ non-overlapping areas}\label{partitioning}
\end{figure}

Fig. \ref{partitioning} shows the sketch of a decomposed power system along with three types of buses for area 1. The state variables of these three types of bus, for $i = 1, 2, \cdots, r$, are represented by ${x_i}^\text{int}$, ${x_i}^\text{bnd}$ and ${x_i}^\text{ext}$ that contain the voltage magnitude and phase angle of internal buses, boundary buses, and external buses, respectively.

Consequently, The state vector of $i$th area is introduced as follows
\begin{align}
x_i = [{x_i}^\text{int}, {x_i}^\text{bnd}, {x_i}^\text{ext}]^T
\end{align}

The dimension of state vector $x$ at area $i$ is $n_i$. Under the above definition, the state vector of each area contains not only its variables, but also parts of the state variables of other areas. Therefore, some state variables will be estimated simultaneously by several neighboring estimators. 

\begin{figure}
  \centering
  \includegraphics[scale=0.55]{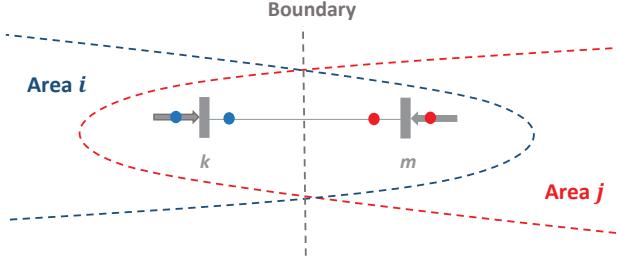}
  \caption{Boundary measurements in decomposed power network}\label{boundary}
\end{figure}

By decomposition of a power network into separate areas, every subsystem has its own measurements and also there would be some boundary measurements. Fig. \ref{boundary} shows the boundary measurements in buses $m$ and $k$ in two different areas. In fact, the boundary between two areas determines which measurement belongs to which area. As illustrated in Fig. \ref{boundary}, the power injection to bus $k$ and power flow from this bus at area $i$, cannot be available at area $j$ and the same concept is established for area $j$. Hence, the boundaries are very deciding in each subsystems' measurements.

Each area has a reference bus with an arbitrary value and therefore, other phase angles are determined with respect to this reference bus. One bus at area 1 is usually chosen as the global reference bus for the entire system. Note that the voltage phase angle of the reference buses will not be estimated unless they have been equipped with PMUs.

\section{Least Square Based SE}\label{section3}
In this section, we concentrate on the basics of the least square based estimation that is used in the proposed approach. 
Consider a nonlinear model as follows
\begin{equation}\label{6}
z=h(x)+e=\
\begin{bmatrix}
    z_1 \\
    z_2 \\
    \vdots \\
    z_m
  \end{bmatrix}=\
  \begin{bmatrix}
    h_1(x_1,x_2,...,x_n) \\
    h_2(x_1,x_2,...,x_n) \\
    \vdots \\
    h_m(x_1,x_2,...,x_n)
  \end{bmatrix}+\
    \begin{bmatrix}
    e_1 \\
    e_2 \\
    \vdots \\
    e_m
  \end{bmatrix}
\end{equation}
where $z$ is the measurement vector, $x$ is the unknown state variable vector, $h$ is the vector of nonlinear functions that represent the relationship between the measurements and the states of the system, and $e$ is the measurement error vector that indicates the mismatch between the measured vector $z$, and the vector $h(x)$. It is assumed that the measurement errors are independent variables, with zero mean and a diagonal covariance matrix as follows
\begin{equation}\label{2}
W=\text{cov}(e)=\mathbb{E}(e.e^T)=\diag\{\sigma_1^2,\sigma_2^2, ... ,\sigma_m^2\}
\end{equation}
where $\sigma_i$ is the standard deviation of errors and $i$ refers to the measurement number.

The WLS method tries to estimate the state variables by minimizing the following performance index.
\begin{equation}\label{7}
J(x)=(z-h(x))^T W^{-1} (z-h(x))
\end{equation}
where $W\in \mathbb{R}^{m \times m}$ is a weighting matrix.

By applying a first-order Taylor series expansion and using the Gauss-Newton iterative solution for \eqref{7}, the SE can be calculated from the following recursive equation
\begin{equation}\label{9}
x^{k+1}=x^{k}+\Delta x^{k}
\end{equation}

The iterations will continue until the alteration in state variables has fallen within a satisfactory range as
\begin{equation}\label{10}
\max ~\| \Delta x^{k} \| < \epsilon
\end{equation}
where $\epsilon$ is a small positive number and indicates the accuracy of the estimation.

By defining $H(x)=\frac{\partial h(x)}{\partial x}$, the covariance matrix of these estimated variable can be achieved as follows
\begin{equation}\label{11}
\text{cov}(\hat{x})=(H^T(\hat{x})W^{-1}H(\hat{x}))^{-1}
\end{equation}

\subsection{Robust SE}\label{Robust}
Although power systems are nonlinear, with some approximations they can be modeled as a linear system. In the following, we investigate a robust estimation based on a linear model as follows
\begin{equation}\label{1}
z=Hx+e
\end{equation}
The above-mentioned model is a linear version of \eqref{6} in which the Jacobian matrix $H$ includes all derivatives of measurement equations with respect to state variables.
The nominal data $\{z,H\}$ in \eqref{1} are often subjected to disturbances and uncertainties. Such errors can deteriorate the performance of the estimation process. To solve this issue, the robust least squares problem is presented as a min--max optimization that is known as bounded data uncertainty (BDU) problem \cite{r36}. The BDU problem is trying to obtain an estimation that minimizes the worst case possible of the cost function in presence of uncertainty.
Consider the model presented in \eqref{1} subject to uncertainties as follows
\begin{equation}\label{12}
(z+\delta z)=(H+\delta H)x+e
\end{equation}
where $\delta z$ denotes an $m\times 1$ perturbation vector, and $\delta H$  is an $m\times n$ perturbation matrix. Note that these quantities are considered as unknown bounded values which are different from $\sigma$, the standard deviation of meters.

In this model, both the measurements and the network parameters are assumed to be uncertain with the following structure
\begin{equation}\label{13}
[\begin{matrix}\delta H & \delta z\end{matrix}]=\
S\Delta[\begin{matrix}E_h & E_z\end{matrix}]
\end{equation}
where $S$, $E_h$ and $E_z$ are known parameters with appropriate dimensions, and $\Delta$ represents an unknown bounded uncertainty that satisfies
\begin{equation}\label{14}
\| \Delta \| \leq1
\end{equation}

\begin{remark}
Perturbation model like \eqref{13} is common in robust estimation and control. It could arise from tolerance specifications on physical parameters \cite{cheng}.
\end{remark}

Considering the model in \eqref{12}, the cost function given in \eqref{7} is changed to the following robustified version
\begin{equation}\label{16}
J(x,y)=\min_{x}\max_{\parallel y \parallel \leq \phi(x)}[(Hx-z+Sy)^T R(Hx-z+Sy)]
\end{equation}
where $R=W^{-1}$ is the weighting matrix.

Here, $y$ is a $q\times 1$ unknown perturbation vector and $S$ is an $m\times q$ known matrix, where $q$ is the dimension of uncertainty parameter $\Delta$. Since $y$ itself is not known as a set of uncertainty, we consider a bound on its Euclidean norm. The upper bound of $y$  is ${\textstyle{\phi(x)=\| E_h x-E_z \|}}$, that is a nonnegative function of $x$. We can also define
\begin{equation}\label{17}
Sy \triangleq{\delta Hx-\delta z}
\end{equation}
The estimated values of \eqref{16} can be obtained as
\begin{equation}\label{18}
\hat{x}=\arg \min_{x}\max_{\parallel y \parallel \leq \phi(x)} J(x,y)
\end{equation}

Consequently, we have a constrained two-player game problem which needs to be solved once for any fixed $x$ and once for any fixed $y$. The solution of this problem is achieved in \cite{r34} as follows
\begin{equation}\label{19}
\hat{x}=(H^T\hat{R}H)^{-1}{(H^T\hat{R}z+\hat{\lambda}E_h^T E_z)}
\end{equation}
where the modified weighting matrix $\hat{R}$ are derived from $R$ via
\begin{equation}\label{20}
\hat{R}\triangleq{R+RS(\hat{\lambda}I-S^T RS)^\dag S^T R}
\end{equation}

The nonnegative scalar parameter $\hat{\lambda}$ is obtained from the following optimization problem
\begin{argmini}
{{\scriptstyle\lambda \geq \| S^T R S \|}}{G(\lambda)}
{\label{21}}{\hat{\lambda}=}
{}
\end{argmini}
where the function $G(\lambda)$ is defined as \eqref{22}, in which ${\textstyle{\| Hx(\lambda)-z\|}_{\scriptscriptstyle{R(\lambda)}}}$ is the weighted norm of vector ${\textstyle(Hx-z)}$.
\begin{equation}\label{22}
G(\lambda)=\lambda\| E_h x(\lambda)-E_z \|+{\| Hx(\lambda)-z \|}_{R(\lambda)}
\end{equation}
where
\begin{equation}\label{23}
R(\lambda) = R+RS(\lambda I-S^T RS)^\dag,
\end{equation}
and
\begin{equation}\label{24}
x(\lambda)\triangleq{(H^T R(\lambda)H)^{-1} (H^T R(\lambda)z+\lambda E_h^T E_z )}.
\end{equation}

We thus see that solving \eqref{18} requires an optimal nonnegative scalar parameter $\hat{\lambda}$, and it needs to minimize $G(\lambda)$ over the semi-open interval ${[\parallel S^T R S \parallel,\infty)}$. Despite the difficulties of solving the optimization problem \eqref{21}, it has a unique solution \cite{r34}. Furthermore, in \cite{r35} 
a practical approximation for $\lambda$ is given as
\begin{equation}\label{25}
\lambda=(1+\mu)\| S^T R S\|
\end{equation}
where $\mu\in \mathbb{R}^+$ is an arbitrary value to attain a suitable performance.

\section{Framework for Multi-Area Robust SE} \label{framework}
In this section, we first present the robust hybrid SE to discuss using both SCADA and PMU measurements in SE, and then the proposed two-level robust SE approach  for multi-are power systems is investigated.

\subsection{Robust Hybrid SE}\label{Hybrid}
In the proposed method, the hybrid SE consists of two steps in which traditional and PMU measurements are processed in individual state estimators, and then joined in a linear one. By this way, the difference between SCADA and PMU's sampling rate will not be problematic; furthermore, there is no need to change the SCADA-based estimators to incorporate PMU measurements as they are employed in separate estimators. At the first step, we estimate the states using the traditional SCADA measurements via the nonlinear model described at the beginning of the Section \ref{section3}. At the second step, we solve a robust SE problem using phasor measurements along with the results of the first step. This combination leads to more accurate estimations. The second step's robust linear SE is discussed in Section \ref{Robust}. Note that one PMU can measure not only the voltage phasor, but also the line's current phasors. Therefore, a PMU can give more than one measurement.

\begin{remark}
The results of the first step of estimation are in polar coordinates. Prior to be used at the second step, the obtained state vector of the first step should be transformed from polar to rectangular coordinates for the linearity requirement to be satisfied.
\end{remark}

The complete measurement model is shown in \eqref{26}, in which $z$ is the vector containing both traditional and PMU measurements and its elements are in rectangular coordinates. $V_R$, $I_R$, $V_I$ and $I_I$ are the real and imaginary parts of bus voltage and current, respectively. The subscripts TSE and PMU refer to values obtained from the traditional SE and PMU measurements, respectively. Since the state variables are the real and imaginary parts of voltage, the partial derivatives in Jacobian matrix $H$ are with respect to $V_R$ and $V_I$. Also, $e$ is the measurement error vector in which related elements to TSE are obtained from \eqref{11}. Likewise, since all the amounts in this technigue are indicated in rectangular coordinates, the standard deviations of variables should also be converted to rectangular format.

\begin{equation}\label{26}
\begin{scriptsize}
\begin{multlined}
z\!=\!
\begin{bmatrix}
    {\begin{bmatrix}V_R\\V_I\end{bmatrix}}_\text{TSE} \\[7pt]
    {\begin{bmatrix}V_R\\V_I\end{bmatrix}}_\text{PMU} \\[7pt]
    {\begin{bmatrix}I_R\\I_I\end{bmatrix}}_\text{PMU}
  \end{bmatrix}
  \!=\!
  \begin{bmatrix}
   \frac{\partial V_{R,\text{TSE}}}{\partial V_R}&\frac{\partial V_{R,\text{TSE}}}{\partial V_I} \\[5pt]
    \frac{\partial V_{I,\text{TSE}}}{\partial V_R}&\frac{\partial V_{I,\text{TSE}}}{\partial V_I} \\[5pt]
    \frac{\partial V_{R,\text{PMU}}}{\partial V_R}&\frac{\partial V_{R,\text{PMU}}}{\partial V_I} \\[5pt]
    \frac{\partial V_{I,\text{PMU}}}{\partial V_R}&\frac{\partial V_{I,\text{PMU}}}{\partial V_I} \\[5pt]
    \frac{\partial I_{R,\text{PMU}}}{\partial V_R}&\frac{\partial I_{R,\text{PMU}}}{\partial V_I} \\[5pt]
    \frac{\partial I_{I,\text{PMU}}}{\partial V_R}&\frac{\partial I_{I,\text{PMU}}}{\partial V_I}
  \end{bmatrix}\!.\!
  \begin{bmatrix}
    V_R \\[5pt]
    V_I
  \end{bmatrix}
  \!+\!
    \begin{bmatrix}
    e_{V_{R,\text{TSE}}} \\[4pt]
    e_{V_{I,\text{TSE}}} \\[4pt]
    e_{V_{R,\text{PMU}}} \\[4pt]
    e_{V_{I,\text{PMU}}} \\[4pt]
    e_{I_{R,\text{PMU}}} \\[4pt]
    e_{I_{I,\text{PMU}}}
  \end{bmatrix}
  \end{multlined}
  \end{scriptsize}
\end{equation}

The measurement Jacobian coefficient matrix $H$, can be rewritten in the following form
\begin{equation}\label{27}
H=\
  \begin{bmatrix}
     \textrm{I} & 0 \\[5pt]
      0 & \textrm{I} \\[5pt]
    \textrm{II} & 0 \\[5pt]
    0 & \textrm{II} \\[5pt]
    \frac{\partial I_{R,\text{PMU}}}{\partial V_R}&\frac{\partial I_{R,\text{PMU}}}{\partial V_I} \\[5pt]
    \frac{\partial I_{I,\text{PMU}}}{\partial V_R}&\frac{\partial I_{I,\text{PMU}}}{\partial V_I}
  \end{bmatrix}
\end{equation}
where \textrm{I} is an identity matrix of dimension $n\times n$ because TSE measurements are directly related to state variables, and \textrm{II} denotes a $p\times n$ sparse matrix which $p$ refers to the number of PMUs. Each element of this matrix is zero except those are related to the buses which have a PMU. In other words, depending on the placement of PMUs, the zero elements of this matrix are replaced by 1. In addition, the covariance matrix $W$, has been expanded to include the standard deviations of the TSE and PMU measurement values as follows

\begin{equation}\label{28}
\begingroup 
\setlength\arraycolsep{0.1pt}
\begin{scriptsize}
W_h=
  \begin{bmatrix}
    {{\sigma^{{\scriptscriptstyle 2}}}_{{\tiny V_{R,\text{TSE}}}}} & 0 & 0 & 0 & 0 & 0\\
    0 & {{\sigma^{\scriptscriptstyle 2}}_{{\tiny V_{I,\text{TSE}}}}} & 0 & 0 & 0 & 0\\
    0 & 0 & {{\sigma^{\scriptscriptstyle 2}}_{{\tiny V_{R,\text{PMU}}}}} & 0 & 0 & 0\\
    0 & 0 & 0 &{{\sigma^{\scriptscriptstyle 2}}_{{\tiny V_{I,\text{PMU}}}}} &  0 & 0\\
    0 & 0 & 0 & 0 & {{\sigma^{\scriptscriptstyle 2}}_{{\tiny I_{R,\text{PMU}}}}} &  0\\
    0 & 0 & 0 & 0 & 0 & {{\sigma^{\scriptscriptstyle 2}}_{{\scriptscriptstyle I_{I,\text{PMU}}}}}\\
  \end{bmatrix}
 \end{scriptsize}
  \endgroup
\end{equation}

In \eqref{28}, each element, $\sigma^2$, is a diagonal matrix containing the individual variances within the distinct measurement type. Now, to solve the hybrid estimation problem, we can apply a cost function as follows 
\begin{equation}\label{extra}
J_\text{hybrid}(x)=(z-Hx)^T W_h^{-1} (z-Hx)
\end{equation}
Therefore, the weighted least squares solution of this model is non-iterative and is given by
\begin{equation}\label{29}
\hat{x}_\text{hybrid}=(H^T W_h^{-1} H)^{-1} H^T W_h^{-1} z
\end{equation}

This two-step estimation leads to faster process due to less dimension, and no need to run further iterations at the second step, as well as more accurate results owing to inclusion of high-precision PMU measurements. The linearity of the second step is the motivation of consolidation of structured uncertainty \eqref{13} into existing hybrid method. In the following subsection, we will propose a framework for multi-area \textit{robust} SE by adding uncertainty into the model of the decomposed power system. In fact, we will solve a distributed hybrid SE with this difference that the matrices i.e. the network parameters and the measurements, are implicitly uncertain.

\begin{remark}
Hybrid estimation is used so as to make the estimation process linear to pave the way for utilizing structured uncertainty which has not been applied to power system so far. In view of the fact that linearization of power system has difficulties and also needs the operating point which is unknown, using two-step hybrid estimation comes across as the best way.
\end{remark}

\subsection{Proposed Two-level Robust SE for Multi-Area Power Systems}
In this section, we propose a framework for multi-area robust SE in a large-scale power system in which both measurements and network parameters are implicitly uncertain. The main goal is to find a robust solution for voltage phasors in all buses.
The proposed multi-area approach consists of the two following levels. 

\subsubsection{Level 1 -- Local Area SE}\label{stage1}
At the first level, each subsystem uses a robust hybrid estimator to estimate its own state variables by minimizing the following cost function as the first step of hybrid technique.
\begin{equation}\label{30}
\min_{z_i=h_i(x_i)+e_i} J_i = {e_i}^T {W_i}^{-1} e_i
\end{equation}
where
\begin{itemize}
	\item $z_i$ is the measurement vector at area $i$ which has $m_i$ elements. It includes not only internal measurements but also boundary measurements such as power injections and power flows in boundary buses of area $i$.
	\item $e_i$ is the measurement error vector at area $i$.
	\item $W_i$ is the covariance matrix at area $i$
    \item $h_i(x_i)$ is the nonlinear measurement function at area $i$.
\end{itemize}

The results of the first level should be \textit{unbaised} estimations of internal states of each area, $\hat{x}_i^\text{int},~ (i=1,\cdots,r)$. 
Otherwise, it is required to augment the number of meters or revise their placement. 

\subsubsection{Level 2 -- Central Coordinator}\label{stage2}
At the second level, the central coordinator tries to obtain an accurate estimation of the boundary state variables by processing the data received from all the individual areas along with the boundary and PMU measurements. Therefore, the following optimization problem is considered

\begin{equation}\label{31}
\min_{z_c=h_i(x_c)+e_c} J_c = {e_c}^T {W_c}^{-1} e_c
\end{equation}
where ${\textstyle{z_c=[z_b^T, z_{\scriptstyle \text{PMU}}^T, \hat{x}^{\text{bnd}^T}, \hat{x}^{\text{ext}^T}]^T}}$ is the measurement vector of the central coordinator, $z_b$ is the boundary measurement vector including power injections in boundary buses and power flows in boundary tie-lines, $z_{\scriptstyle \text{PMU}}$ is the PMU measurement vector, ${\textstyle{\hat{x}^\text{bnd}=[\hat{x}_1^{\text{bnd}^T}, \hat{x}_2^{\text{bnd}^T}, ..., \hat{x}_r^{\text{bnd}^T}]^T}}$ and ${\textstyle{\hat{x}^{\text{ext}}=[\hat{x}_1^{{\text{ext}}^T}, \hat{x}_2^{{\text{ext}}^T}, ..., \hat{x}_r^{\text{ext}^T}]^T}}$ are the estimated boundary and external state variables vector by local estimators, respectively. These estimated values are considered as pseudo-measurements in the central coordinator because they have priorly been estimated at level 1 by local estimators. 
The cost function \eqref{31} tries to estimate ${\textstyle{x_c=[x^{\text{bnd}^T}, u^T]^T}}$ 
in which $u$ is the reference bus vector with respect to the global reference bus at area 1. 



\begin{figure}[!h]
  \centering
  \includegraphics[scale=0.52]{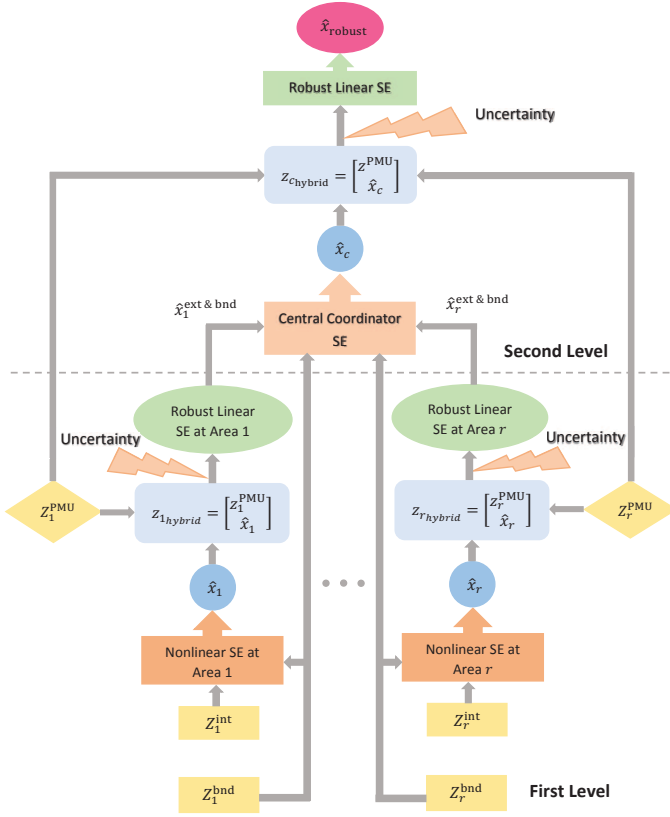}
  \caption{The structure of the proposed multi-area robust SE in power system}\label{structure}
\end{figure}

As a whole, at the first level, we use a robust hybrid technique at each area, then at the second level by using the results of the previous level as pseudo-measurements and other available boundary and PMU measurements, a global robust hybrid SE is performed by the central coordinator. Note that every estimator uses a hybrid-based model in estimation process as introduced in Section \ref{Hybrid}. The structure of the method is shown in Fig. \ref{structure}. Moreover, details of the proposed multi-area robust estimation approach is presented in Algorithm \ref{alg}.

\begin{remark}
The advantages of the proposed multi-area hybrid estimation can be listed as follows:
\begin{itemize}
  \item Increasing estimation accuracy by incorporating both unsynchronized traditional and synchronized phasor measurements.
  \item No requirements for changing the existing SCADA software package owing to utilizing a non-invasive manner.
  \item The difference between the sampling rates of SCADA and PMU measurements will not be troublesome.
  \item Reducing computational time due to less dimension, no need for iterations at the second linear step, and the capability of parallel processing at areas at the first level of proposed two-level algorithm.
\end{itemize}
\end{remark}

\begin{remark}
If the model does not have uncertainty, the proposed method will be reduced to a non-robust multi-area hybrid estimation by setting the uncertainty parameters to zero in the algorithm. Furthermore, if the real system does not have uncertainties and we run a robust algorithm for SE, the results will be sub-optimal because of the conservatism of the robust methods.
\end{remark}

\begin{algorithm}[!h]
  \caption{Multi-area robust SE in a power system}
  \label{alg}
   \begin{center}
  \textbf{Level 1}
   \end{center}
   \small
Decompose the power system into $r$ non-overlapping areas and do the following items for each area in parallel.

  \textsf{Initial condition:} Set $x_0=0$ rad, for the voltage phase angle, and $x_0=1$ p.u, for the voltage magnitude at all buses. Set $k = 0$ as the iteration index

   \begin{enumerate}
  \item Input $z_{{\scriptstyle \text{RTU}}}$ , $W$, $k_{\scriptstyle \text{limit}}$, $\epsilon$
  \item Determine $\hat{x}_i$ by minimizing $J(x)$ in \eqref{30}
  \item Convert the estimates and estimation errors obtained from \eqref{9} and \eqref{11}, respectively, from polar coordinates to rectangular coordinates 
  \item Input PMU measurements in rectangular coordinates, $(z_{\scriptstyle \text{PMU}})$
  \item Construct a new hybrid measurement vector,
  $z_{\scriptstyle \text{hybrid}}= (\hat{x}_{r}, ~ \hat{x}_{i}, ~ z_{\scriptstyle \text{PMU}})^T$ \\
  \item Construct $H_{\scriptstyle \text{hybrid}}$ as in \eqref{27}
  \item Add uncertainty to $z_{\scriptstyle \text{hybrid}}$ and $H_{\scriptstyle \text{hybrid}}$ \\
  ${( \textstyle{H_{\scriptstyle \text{hybrid}} \longrightarrow H_{\scriptstyle \text{hybrid}}+\delta_{H_{\scriptscriptstyle \text{hybrid}}} \hskip 0.5em , \hskip 0.5em z_{\scriptscriptstyle \text{hybrid}} \longrightarrow z_{\scriptscriptstyle \text{hybrid}}+\delta_{z_{\scriptscriptstyle \text{hybrid}}}}) }$ 
  \item For the uncertain model \eqref{12}--\eqref{13}, choose an appropriate value for $\lambda$  as in \eqref{25}
  \item Determine $\hat{x}_{\scriptstyle \text{robust}}$ by solving the min-max optimization problem in \eqref{16}
  \item Convert the estimates from rectangular coordinates to polar coordinates
  \end{enumerate}

     \begin{center}
  \textbf{Level 2}
   \end{center}
  \begin{enumerate}
  \item Receive the pseudo-measurements from the previous level $(\hat{x}_\text{bnd})$
  \item Input boundary measurements from each area $(z_\text{bnd})$
  \item Construct the central coordinator measurement vector 
  ${(z_\text{coordinator}=\begin{bmatrix}\hat{x}_\text{bnd}\\z_\text{bnd}\end{bmatrix})}$ \\[1.5pt]
  \item Determine $\hat{x}_c$ by minimizing $J(x)$ in \eqref{31}
  \item Convert the estimates from rectangular coordinates to polar coordinates
   \item Input PMU measurements in rectangular coordinates, $(z_{\scriptstyle \text{PMU}})$
   \item Construct the central coordinator hybrid measurement vector ${\textstyle{(z_{c_\text{hybrid}}=\begin{bmatrix}\hat{x}_{c_\text{real}}\\\hat{x}_{c_\text{imag}}\\z_{c_\text{PMU}}\end{bmatrix})}}$ \\[1.5pt]
   \item Construct $H_{c_\text{hybrid}}$ as in \eqref{27}
   \item Add uncertainty to $z_{c_\text{hybrid}}$ and $H_{c_\text{hybrid}}$ \\
  ${( \textstyle{H_{c_\text{hybrid}} \longrightarrow H_{c_\text{hybrid}}+\delta_{H_{c_\text{hybrid}}} \hskip 0.2em , \hskip 0.2em z_{c_\text{hybrid}} \longrightarrow z_{c_\text{hybrid}}+\delta_{z_{c_\text{hybrid}}}}) }$
  \item Choose an appropriate value for $\lambda$  as in \eqref{25}
  \item Determine $\hat{x}_{c_\text{robust}}$ by solving the min-max optimization problem in \eqref{16} for the central coordinator
  \item Convert the estimates from rectangular coordinates to polar coordinates
\end{enumerate}
\end{algorithm}
\normalsize
\section{Simulation Results} \label{simulation}
\begin{figure}[!t]
  \centering
  \includegraphics[width=3.5in]{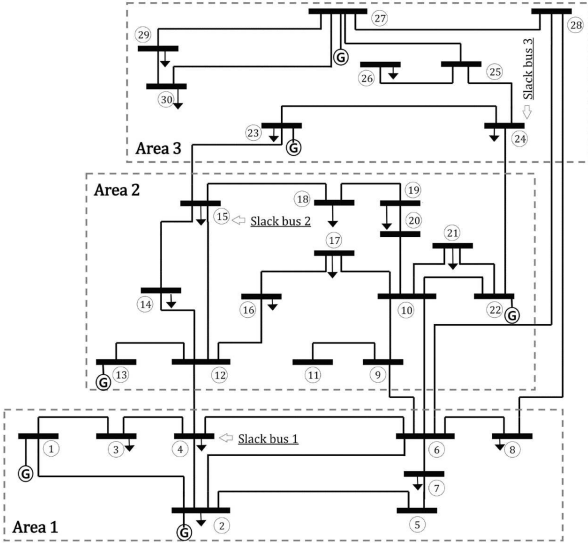}
  \caption{IEEE 30--bus test system diagram}\label{IEEE30}
\end{figure}

To evaluate the performance of the proposed multi-area robust estimation approach, the simulations are carried out with MATLAB software for the IEEE 30--bus test system. The diagram of this network is depicted in Fig. \ref{IEEE30}. As shown, the network has been decomposed into 3 non-overlapping areas which are connected together by multiple tie-lines. In this network, buses 4, 6, 9, 10, 12, 15, 22, 23, 24 and 28 are boundary buses. Buses 4, 15 and 24 are chosen as slack buses at each area and equipped with PMUs. Detailed information about the type and number of buses at each area are presented in Table \ref{table1}.
\begin{table}
\centering
\caption{IEEE 30--bus system area composition}
\scalebox{1}{
\begin{tabular}{c|c c c}
 Area & 1 & 2 & 3 \\
\hline
 Internal buses & 5 & 9 & 5 \\
 Boundary buses & 3 & 5 & 3 \\
 External buses & 4 & 4 & 4 \\
\end{tabular}}
\label{table1}
\end{table}

To produce realistic noisy measurements, a load flow is first ran to generate true data, then the measurement error variance, $\sigma^2$, corresponding to each measurement type is added to them since the existing meters are not accurate. By defining enough measurements, we intend to estimate the unknown variables of this network, including the magnitude and phase angle of all buses. The quantities for standard deviations of measurements from \cite{scada,pmu} are tabulated in Table \ref{table2}. Note that a PMU has a smaller error deviation than other conventional measurements.

\begin{table}[h!]
\centering
\caption{Standard Deviations of the Measurements}
\scalebox{1}{%
\begin{tabular}{c|c|c}
 Power injection & Power flow & PMU \\
\hline
 0.01 & 0.008 & 0.001 \\
\end{tabular}}
\label{table2}
\end{table}

\begin{table}
\centering
\caption{Measurement type and numbers in IEEE 30--bus system}
\scalebox{1}{%
\begin{tabular}{c|c c c}
 Area & 1 & 2 & 3 \\
\hline
 Power injection pairs & 3 & 5 & 3 \\
 Power flow pairs & 15 & 21 & 12 \\
\end{tabular}}\label{table3}
\end{table}

\begin{figure}[!h]
  \centering
  \includegraphics[width=3.5in]{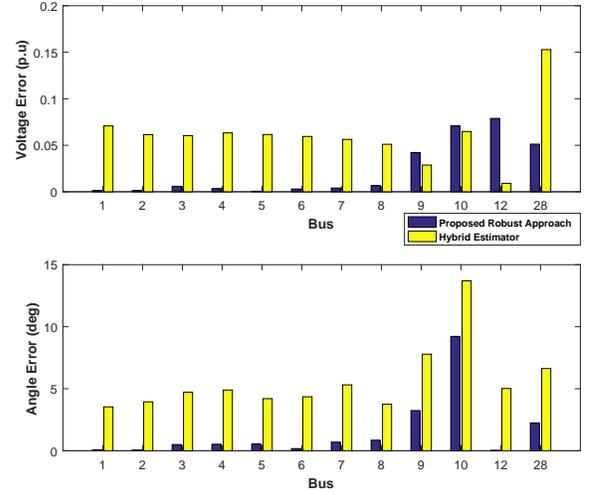}
  \caption{Estimation errors of voltage magnitude and phase angle at area 1}\label{fig1}
\end{figure}

\begin{figure}[!h]
  \centering
  \includegraphics[width=3.5in]{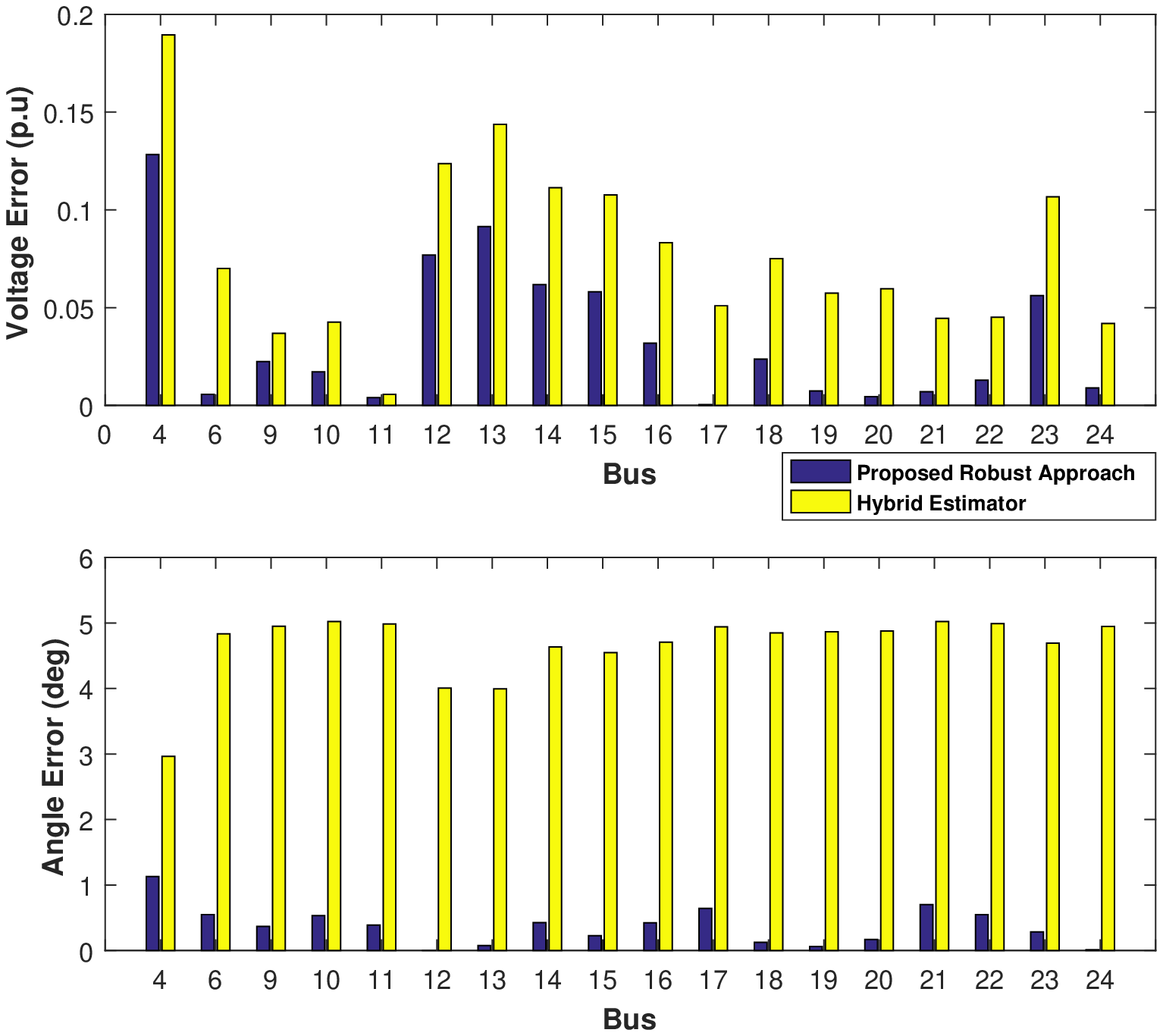}
  \caption{Estimation errors of voltage magnitude and phase angle at area 2}\label{fig2}
\end{figure}

\begin{figure}[!h]
  \centering
  \includegraphics[width=3.5in]{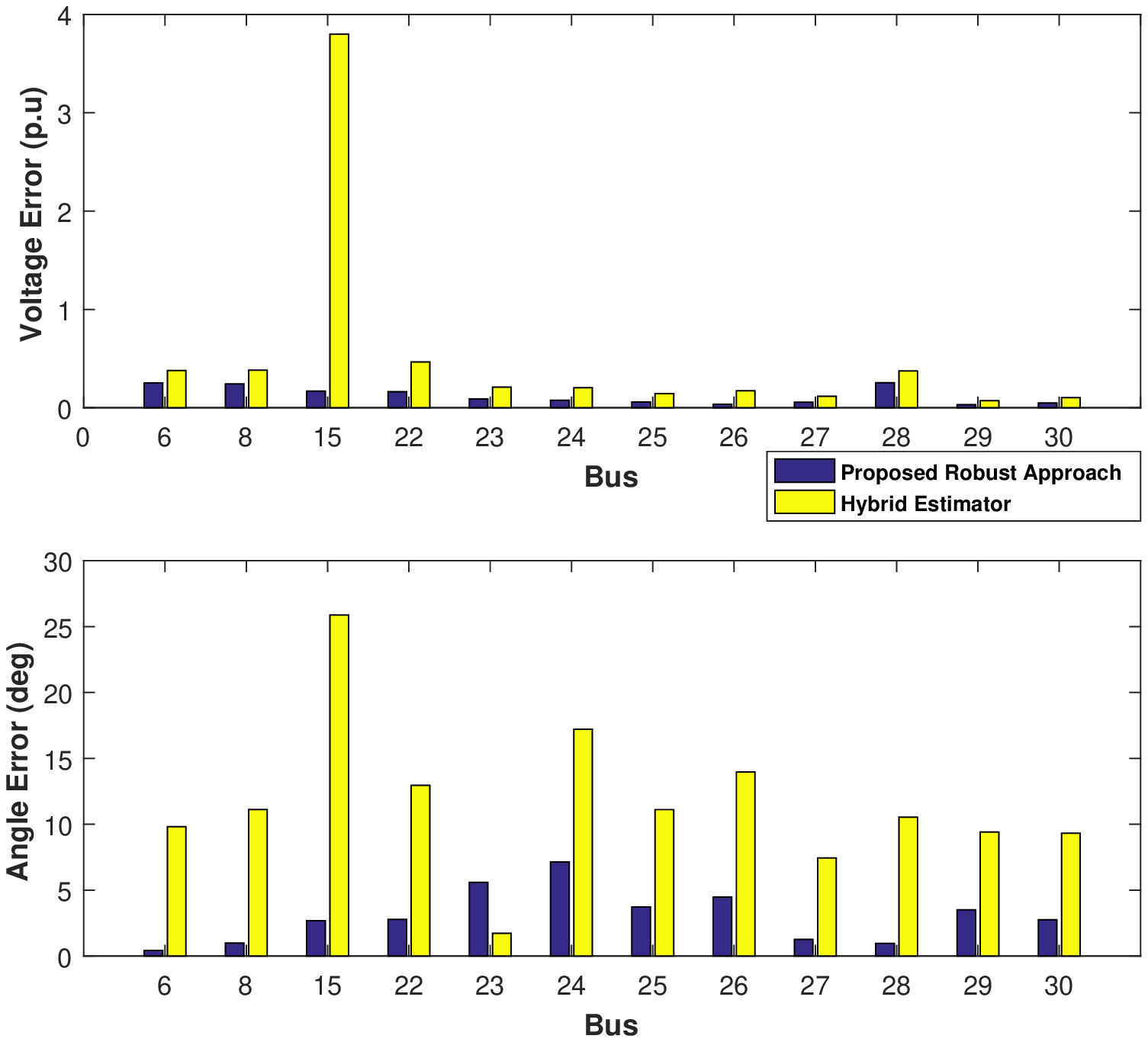}
  \caption{Estimation errors of voltage magnitude and phase angle at area 3}\label{fig3}
\end{figure}

\begin{figure}[!h]
  \centering
  \includegraphics[width=3.5in]{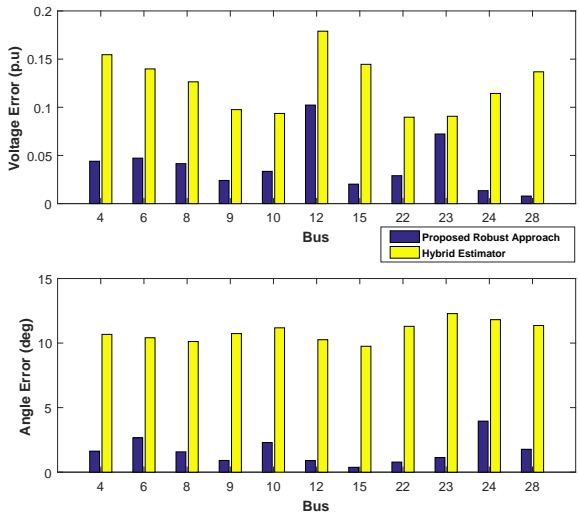}
  \caption{Estimation errors of voltage magnitude and phase angle by the central coordinator}\label{fig4}
\end{figure}

It should be pointed out that in the network decomposition, every area must be observable. Let us define $m$ as the number of measurements, $n$ as the number of variables, and $\eta$ as the ratio of the number of measurements per the number of variables. In this test, we consider this ratio as 1.3. Table \ref{table3} has more detailed information regarding the measurement numbers for the test. One power injection pairs contains active and reactive parts of that. To clarify the performance of the proposed approach, the results are compared with the two-level version of hybrid estimation presented in Section \ref{Hybrid}. Note that both cases are exposed to uncertainty in measurements and network parameters. Fig. \ref{fig1} shows the estimation errors of voltage magnitude and phase angle at each buses at area 1 for the IEEE 30--bus system with two methods. Figs. \ref{fig2} and \ref{fig3} show the same results at area 2 and 3, respectively. The error denotes the absolute difference between estimated state variables and true values which are obtained from a load flow run.
Given that the slack buses are equipped with PMUs in this network, the phase angle of these buses is also estimated. Each area has to estimate its internal buses correctly; therefore, the inaccuracy in estimated boundary and external buses can be disregarded. These buses will be re-estimated in the central coordinator as shown in Fig. \ref{fig4} where the central coordinator estimates the boundary states.
It can be seen from the results that the proposed multi-area robust SE  has the ability to limit the uncertainty and can provide more accurate results, i.e. less estimation error than the WLS method.

\section{Conclusions}\label{conclusion}
In this paper, a novel algorithm for multi-area robust SE in large-scale power systems with bounded uncertainties is proposed. For this purpose, the power network is first decomposed into several subsystems. Then, a two-level algorithm is proposed to find a robust solution for estimating the unknown states of the system. In the proposed approach, at the first level, a nonlinear SE is performed based on the traditional SCADA measurements and the network topology. After that, by using the PMU results as a new measurement set, a robust hybrid SE is done at each area. At the second level, the central coordinator gives a robust central estimation of the entire system by receiving the results of estimated states of each area and other available boundary and PMU measurements. The privilege of this approach is the ability to do a distributed robust estimation in a large-scale power system while both the measurements and network parameters are subjected to structured bounded data uncertainty. By applying the proposed method on the IEEE 30--bus test system, we showed that it can limit the impact of the uncertainties well and provide a better solution compared with the hybrid method.

\balance
\bibliographystyle{IEEEtran}
\bibliography{ref}

\end{document}